\documentclass[a4paper,11pt]{article}
\usepackage{fullpage}
\usepackage{times}
\usepackage{subfigure}
\usepackage{vmargin}
\usepackage{fancyheadings}
\usepackage[cp1251]{inputenc}
\usepackage[english]{babel}
\usepackage{graphicx}
\usepackage{color}
\usepackage{cite}
\usepackage{amsmath,amssymb,amsthm,enumerate,bbm,color,verbatim,setspace}

\date{}
\title{Entopy-energy inequality for superconducting qutrit}
\author{V.I. Man'ko$^{1,2,3}$ and L.A. Markovich$^{3,4*}$\\
      $^1$P.N. Lebedev Physical Institute, Russian Academy of Sciences,\\
      $^2$National Research Tomsk State University,\\
$^3$Moscow Institute of Physics and Technology,\\
$^4$Institute for information transmission problems, Moscow,\\
$^4{^{*}}$Corresponding author e-mail: kimo1@mail.ru}
\date{}
\begin{document}
\maketitle
\pagenumbering{arabic}
\begin{abstract}\noindent
We compare the entropy-energy inequality and the von Neumann entropic inequality for three level atom implemented on superconducting circuits with Josephson junction. The positivity of entropy and energy relations for the qutrit system are used for verification of state tomography  of qudit systems. The results obtained are valid for generic quantum states (qudits) and are illustrated on the example of the temperature density matrix of the single qutrit state.
\end{abstract}
\footnote{Material of the paper is connected with the talk given by V.I.Man'ko at conference in Tomsk, SFP-2016.}
\medskip

\noindent{\bf Keywords:} qutrit,  superconducting circuits , entropic inequalities, entropy-energy inequality

\section{Introduction}\label{sec:14_1}
The notion of quantum entanglement reflecting the presence of quantum correlations \cite{schredinger:35} is well known for composite systems like two-qubits with spins  $j=1/2$.
Composed systems have correlations between subsystems and the physical meaning of these correlations is rather natural  \cite{Can:2005}. However, in the systems without subsystems, like one qutrit or qudit, the nature of entanglement and correlations is not so obvious. These notions for noncomposite systems have been studied relatively recently in \cite{Manko:2014,Chernega:2014,Mar2}. The existence of the correlations between the subsystems is detected by different types of inequalities, like Bells inequality \cite{Clauser:1969,Lieb:1974,Wehner:2010}, violated for the entangled states  \cite{Cirelson:1980}, or by variety of  entropic and information inequalities written for the density matrices of the system and its subsystems. In \cite{Mar8} it is shown that the latter inequalities can be applied for the systems without subsystems.
\par In \cite{Chernega:2013} the method of qubit portrait is obtained to write the new inequalities for the qudit systems. In \cite{Mar5} the latter method is used to investigate the notion of separability and entanglement for the systems with the spin $j=1$ (single qutrit) and in \cite{Chernega:2013} the new entropic inequality for such system is obtained. In \cite{Figueroa} the inequality for the relative entropy \cite{Nielsen} is rewritten in energetic form, called entropy-energy inequality. Using qubit portrait method it can be easily extended to the case of the noncomposite systems.
\par The aim of the paper is to compare the entropy-energy inequality and the von Neumann entropy inequality written for the system of the single qutrit.
It can be realized as a three-level artificial atom in superconduction circuit based on Josephson junction (see e.g.  \cite{Kiktenko2,Glushkov}). The density matrix of the state of the qutrit can be measured (see e.g. \cite{Shalibo}) by the method of quantum tomography \cite{OVManko} in which the quantum states are associated with fair probabilities \cite{Ibort}.
The new approach of extension of the known quantum characteristics for the composite system on the single qutrit system is demonstrated on the example of the density matrix of the system in thermal equilibrium state.
\par The paper is organized as follows. We remind the notion of the qubit portrait method and the von Neumnann entropic inequity for the single qutrit system.
The entropy-energy inequality is written for the single qutrit system and compared with the von Neumnann entropic inequity. The results are illustrated on the example of the temperature density matrix for the different values of temperature.
\section{Qubit portrait}
Let us have the system of the single qutrit with the density matrix
\begin{eqnarray}\rho={\left(
                                 \begin{array}{ccc}
                                   \rho_{11}& \rho_{12}& \rho_{13}\\
                                   \rho_{21}& \rho_{22}& \rho_{23}\\
                                   \rho_{31}& \rho_{32}& \rho_{33}\\
                                 \end{array}
                               \right),\quad \rho=\rho^{\dagger},\quad Tr\rho=1}\,. \label{14_1}
                               \end{eqnarray}
and its eigenvalues $\{\lambda_i\}_{i=1,2,3}$ are nonnegative.
According to the qubit portrait method the density matrix  \eqref{14_1} is added by the zero raw and column to the matrix  $\widetilde{\rho}$ of a size $4\times4$.
The features of the density matrix \eqref{14_1} does not change but the fours zero eigenvalue is added. Such density matrix can describe the state in four-dimensional Hilbert space $\mathcal{H}$. Hence, two "artificial subsystems" can be introduced and  the reduced density matrices for them can be written as
  \begin{eqnarray}\widetilde{\rho_1}={\left(
                              \begin{array}{cc}
                                \rho_{11}+\rho_{22} & \rho_{13} \\
                               \rho_{31} &  \rho_{33} \\
                              \end{array}
                            \right),\quad
                          \widetilde{\rho_2}=\left(
                              \begin{array}{cc}
                                \rho_{11}+\rho_{33} & \rho_{12} \\
                               \rho_{21} &  \rho_{22} \\
                              \end{array}
                            \right)}\,. \label{14_6}
\end{eqnarray}
The von Neumann entropy for the qutrit and its "artificial subsystems" has the following view
\begin{eqnarray}\label{14_3}S&=&-Tr\widetilde{\rho}\ln\widetilde{\rho}=-\sum\limits_{i=1}^{4}\widetilde{\lambda}_i\ln\widetilde{\lambda}_i,\quad
\widetilde{S_1} = -Tr\widetilde{\rho_1}\ln\widetilde{\rho_1},\quad
\widetilde{S_2} =-Tr\widetilde{\rho_2}\ln\widetilde{\rho_2}.
\end{eqnarray}
Hence, the quantum information for the single qutrit system  can be written as
\begin{eqnarray}I_q=\widetilde{S_{1}} +\widetilde{S_{2}}-\widetilde{S_{12}},\quad I_q\geq 0. \label{14_12}
\end{eqnarray}
Using \eqref{14_3}, the following inequality for \eqref{14_1} can be written \cite{Chernega:2013}
 \begin{eqnarray}\label{14_11}
S&\leq& - Tr\{\left(
                              \begin{array}{cc}
                                \rho_{11}+\rho_{22} & \rho_{13} \\
                               \rho_{31} &  \rho_{33} \\
                              \end{array}
                            \right)\ln\left(
                              \begin{array}{cc}
                                \rho_{11}+\rho_{22} & \rho_{13} \\
                               \rho_{31} &  \rho_{33} \\
                              \end{array}
                            \right)\}\\\nonumber
&-&Tr\{\left(
                              \begin{array}{cc}
                                \rho_{11}+\rho_{33} & \rho_{12} \\
                               \rho_{21} &  \rho_{22} \\
                              \end{array}
                            \right)\ln\left(
                              \begin{array}{cc}
                                \rho_{11}+\rho_{33} & \rho_{12} \\
                               \rho_{21} &  \rho_{22} \\
                              \end{array}
                            \right)\},
\end{eqnarray}
Note, that the latter inequality corresponds to the quantum correlations in the single qutrit system that is the noncomposite system.
\section{Entropy-energy inequality}
As a measure of difference between two quantum states the relative entropy can be taken \cite{Nielsen}. Let us have two quantum systems with the density matrices $\rho$ and $\sigma$. Hence, the entropy of the state $\rho$ relatively to the entropy of the state $\sigma$ is given by the following inequality
\begin{eqnarray}\label{14_5}Tr\left(\rho\ln\rho-\rho\ln\sigma\right)\geq 0.
\end{eqnarray}
\par Let $H$ be the matrix of the hamiltonian of the single qudit system and
  \begin{eqnarray}\label{14_6}\sigma=\frac{e^{ H}}{Tr(e^{H})}, \quad \sigma\geq 0, \quad Tr \sigma=1.
\end{eqnarray}
Using the definition of the von Neumnann entropy $S=-Tr(\rho\ln\rho)$ the inequality \eqref{14_5} is rewritten in energetic form \cite{Figueroa} as
\begin{eqnarray}\label{14_8}S+\langle H\rangle\leq \ln(Tr e^{ H})),
\end{eqnarray}
where $\langle H\rangle=Tr(\rho  H)$ is a mean energy. Let us select the density matrix of the thermal equilibrium state
  \begin{eqnarray*}\sigma(\beta)_H=\frac{e^{- \beta H}}{Tr(e^{ -\beta H})}, \quad \sigma(\beta)_H\geq 0, \quad  Tr \sigma(\beta)_H=1,
\end{eqnarray*}
where $T=1/\beta$ is interpreted as a temperature.
Using the partition function $Z(\beta)=Tr (e^{-\beta H})$ the inequality  \eqref{14_8} can be written as
 \begin{eqnarray}\label{14_9}S+\langle H\rangle\leq\ln Z(\beta=-1).
\end{eqnarray}
\subsection{Single qutrit system}
\begin{figure}[ht]
\begin{center}
\begin{minipage}[ht]{0.70\linewidth}
\includegraphics[width=1\linewidth]{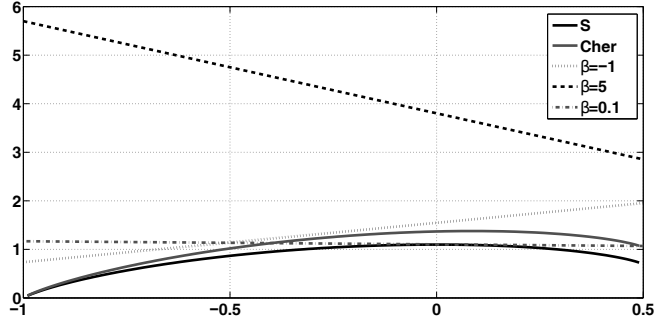}
\vspace{-4mm}
\caption{$S$ (black line)- entropy of the system, $Cher $ (gray line) -  the right hand side of \eqref{14_11}, (dashed lines) - the right hand side o \eqref{14_13} for the different parameters $\beta$.}
\label{fig:14_1}
\end{minipage}
\end{center}
\end{figure}

Let the hamiltonian  $H$ be defined by the following matrix
\begin{eqnarray}H=\left(
                                 \begin{array}{ccc}
                                  H_{11}& H_{12}& H_{13}\\
                                   H_{21}& H_{22}& H_{23}\\
                                   H_{31}& H_{32}& H_{33}\\
                                 \end{array}
                               \right),\quad H=H^{\dagger}. \label{14_10}
                               \end{eqnarray}
Hence, $Tre^{H}= e^{E_1}+e^{E_2}+e^{E_3}$, where $E_1$, $E_2$ and $E_3$ are the energy levels,
$\det(H-EI_3)=0$. Let the elements of the hamiltonian matrix be $H_{12} = H_{21}=H_{23} =H_{32}=0$. Then \eqref{14_8} has the following view
\begin{eqnarray}\label{14_13}S&\leq&-Tr \rho \ln\left(\frac{U\left(
                                 \begin{array}{ccc}
                                   e^{-\beta E_1} & 0& 0 \\
                                   0& e^{-\beta E_2}  &0 \\
                                   0 & 0 & e^{-\beta E_3}  \\
                                 \end{array}
                               \right)
U^{\dagger}}{Tr\{U\left(
                                 \begin{array}{ccc}
                                   e^{-\beta E_1} & 0& 0 \\
                                   0& e^{-\beta E_2}  &0 \\
                                   0 & 0 & e^{-\beta E_3}  \\
                                 \end{array}
                               \right)
U^{\dagger}\}}\right),
\end{eqnarray}
where $U$ is the unitary matrix and its columns are the eigenvectors of  $H$, and its eigenvalues are
 \begin{eqnarray} E_1&=&H_{22},\quad
 E_{2,3}=\frac{H_{11} + H_{33} \mp\sqrt{H_{11}^2 - 2H_{11}H_{33} + H_{33}^2 + 4H_{13}H_{31}}}{2}.
\end{eqnarray}
It is obvious, that the choice of the hamiltonian and the temperature parameter  $\beta$ define the relation between the two inequalities.
\\When the temperature parameter $\beta=0$ the inequality \eqref{14_13} is $S\leq\ln3$ and when $\beta\rightarrow\infty$ the right hand side of the inequality  \eqref{14_13} tends to infinity. As an example let us select the following state and the hamiltonian
\begin{eqnarray}\rho=\frac{1}{3}\left(
                                 \begin{array}{ccc}
                                   1+b& 0& 0\\
                                   0& 1+b& 0\\
                                   0& 0& 1-2b\\
                                 \end{array}
                               \right), H=\left(
                                 \begin{array}{ccc}
                                   1& 0& 1\\
                                   0& -1& 0\\
                                   1& 0& 1\\
                                 \end{array}
                               \right),\label{14_16}
                               \end{eqnarray}
where $-1\leq b\leq1/2$, hold. The results are shown in Fig.~\ref{fig:14_1}.  The entropy $S$ is shown by the black line,  the right hand side of the inequality \eqref{14_11} is shown by the gray line. The dashed lines show the right hand sides of the inequality \eqref{14_13} for the different values of the parameter  $\beta=\{-1;5;0.1\}$.
It is easy to see, that the choice of the parameter $\beta$ determine the value of the right hand side of the energetic inequality. Hence, we can select such parameter that provides the minimum upper bound of the entropy of the system for the energy inequity. 

\subsection{Conclusion}
To conclude let us point out the main results of the paper. We investigate the recently found relations  between the entropy-energy \cite{Figueroa} and the von Neumann entropic inequality \cite{Chernega:2014} for the single qutrit system. These relations and entropic inequalities can be checked experimentally with superconducting circuits where the density matrices of the qutrit states are measured with the quantum tomography method. Note that the single qutrit system is the system without subsystems. The obtained relations can be expressed in terms of tomographic probabilities in explicit form. This provides the new method of studding the quantum correlations in the single qudit systems since there exist a possibility to measure the probabilities directly. The measuring of the probabilities gives information on possible strong quantum correlations in the noncomposite systems. The correlations in this case correspond to analogs of such phenomena as the entanglement characterizing the strong quantum correlations in the multipartite systems.   It is shown that for the thermal state the value of temperature determines the dependence between two inequalities. For the high temperature the right hand side of the energetic inequality exceed  the right had side of the entropic inequality. For the small temperature the right right hand side of the energetic inequality tends to constant. It can be easily concluded that it is possible to find the value of temperature that provides the minimum of the right hand side of the energetic inequality, in other words the minimum upper bound of the entropy of the qudit system for the energy-entropy inequality.

\section*{Acknowledgements}
The research in section 2 by Man'ko V.I. carried out in 2015 was supported by the Tomsk state university competitiveness improvement program.
The research in section 3  by Markovich L.A. was partly supported by the Russian Foundation for Basic Research, grant 16-08-01285 A.


\begin{thebibliography}{8}

\bibitem{schredinger:35} E. Schr{\"{o}}dinger, {\sl Naturwissenschaften}, \textbf{23}, 807--812, 1935.
    \bibitem{Can:2005}M.A. Can, A.A. Klyachko, A.S. Shumovsky,  {\sl J. of Optics B}, \textbf{7(2)}, L1--L3, 2005.
    \bibitem{Manko:2014}M.A. Man'ko, V.I. Man'ko,  {\sl Phys. Scr.}, \textbf{T160}, 014030, 2014.
\bibitem{Chernega:2014}V.N. Chernega, O.V. Man'ko, V.I. Man'ko,  {\sl J. Russ. Laser Res.}, \textbf{35(3)}, 278--290, 2014.
\bibitem{Mar2}V.I. Man'ko, L.A. Markovich, {\sl J. Russ. Laser Res.}, \textbf{35(4)}, 355--361, 2014
\bibitem{Clauser:1969}J.F. Clauser, M.A. Horne, A. Shimony, R.A. Holt, {\sl Phys. Rev. Lett.}, \textbf{23(15)}, 880, 1969.
\bibitem{Lieb:1974}E.H. Lieb, M.B.  Ruskai,  {\sl Adv. Math}, \textbf{12}, 269--273, 1974.
\bibitem{Wehner:2010}S. Wehner, A. Winter, {\sl J. New J. Phys.}, \textbf{12}, 025009, 2010.
\bibitem{Cirelson:1980}B.S. Cirel'son, {\sl J.Lett. Math. Phys.}, \textbf{4(2)}, 93--100, 1980.
\bibitem{Mar8}V.I. Man'ko, L.A. Markovich,  {\sl J. Russ. Laser Res.}, \textbf{35(5)}, 518--524, 2014.
\bibitem{Chernega:2013}V.N. Chernega, O.V. Man'ko, V.I. Man'ko, {\sl J. Russ. Laser Res.}, \textbf{34(4)}, 383--387, 2013.
\bibitem{Mar5}V.I. Man'ko, L.A. Markovich, {\sl J. Russ. Laser Res.},  \textbf{36(2)}, 110--118, 2015.
\bibitem{Figueroa}A. Figueroa, J. Lopez, O. Castanos, R. Lopez-Pena, M.A. Man'ko,  V.I Man'ko, {\sl J. of Phys. A: Math. and Theor.},  \textbf{48(6)}, 065301, 2015.
    \bibitem{Nielsen} M.A. Nielsen, I.L. Chuang, {\sl Cambridge University Press}, 2010.
\bibitem{Kiktenko2}E.O. Kiktenko, A.K. Fedorov, A.A. Strakhov, V.I. Man'ko,    {\sl Phys. Lett. A},  \textbf{379(22)}, 1409--1413, 2015.
\bibitem{Glushkov}A. Glushkova, E. Glushkov, V.I. Manko, {\sl J. Russ. Laser Res.}, \textbf{37(3)}, 236--243, 2016.
\bibitem{Shalibo}Y. Shalibo, R. Resh, O. Fogel, D. Shwa, R. Bialczak, J.M. Martinis, N. Katz, {\sl Phys. Rev. Lett.},  \textbf{110}, 100404, 2013.
 \bibitem{OVManko}V.I. Man'ko,  O.V. Man'ko, {\sl JETP}, \textbf{85 (3)}, 430, 1997. 
     \bibitem{Ibort} A. Ibort,  V.I. Man'ko, G. Marmo, A. Simoni, F. Ventriglia, {\sl Phys. Scr.} \textbf{79(6)}, 2009.


\end{thebibliography}
\end{document}